\documentclass[12pt]{article}

\usepackage{latexsym,amsmath,amssymb,pstricks,wasysym,stmaryrd,enumerate,
epsfig}

\newcommand{\be}{\begin{equation}}
\newcommand{\ee}{\end{equation}}
\newcommand{\beu}{\begin{equation*}}
\newcommand{\eeu}{\end{equation*}}
\newcommand{\bea}{\begin{eqnarray}}
\newcommand{\eea}{\end{eqnarray}}
\newcommand{\beaa}{\begin{eqnarray*}}
\newcommand{\eeaa}{\end{eqnarray*}}
\newcommand{\bmx}{\begin{pmatrix}}
\newcommand{\emx}{\end{pmatrix}}

\newcommand{\del}{\partial}
\newcommand{\g}{{\frak g}}
\newcommand{\h}{{\frak h}}
\newcommand{\m}{{\frak m}}

\newcommand{\nn}{\nonumber}

\newcommand{\eps}{\epsilon}
\newcommand{\tr}{{\rm Tr}}

\advance\textheight by \topskip
\begin{document}
\baselineskip 17pt
\parindent 8pt
\parskip 9pt

\begin{flushright}
DAMTP-2004-33
\break
hep-th/0404003v2\\[3mm]
\end{flushright}

\begin{center}
{\Large {\bf Non-local charges and quantum integrability of }}\\[4mm] 
{\Large {\bf sigma models on the symmetric spaces}}\\[4mm]
{\Large {\bf $SO(2n)/SO(n){\times}SO(n)$ and $Sp(2n)/Sp(n){\times}Sp(n)$ }}\\
\vspace{0.8cm} 
{\large J.M. Evans,
%\footnote{\tt J.M.Evans@damtp.cam.ac.uk}, 
D. Kagan,
%\footnote{\tt D.Kagan@damtp.cam.ac.uk}, 
C.A.S. Young,
%\footnote{\tt C.A.S.Young@damtp.cam.ac.uk}}
}
\\
{\em DAMTP, Centre for Mathematical Sciences, University of
Cambridge,\\ Wilberforce Road, Cambridge CB3 0WA, UK}
\\
{\small E-mail: {\tt J.M.Evans@damtp.cam.ac.uk}, 
{\tt D.Kagan@damtp.cam.ac.uk}, {\tt C.A.S.Young@damtp.cam.ac.uk}}
\\

\end{center}

\vskip 0.15in
 \centerline{\small\bf ABSTRACT}
\centerline{
\parbox[t]{5in}{\small
\noindent 
Non-local conserved charges in two-dimensional sigma models 
with target spaces 
$SO(2n)/SO(n){\times}SO(n)$ and 
$Sp(2n)/Sp(n){\times}Sp(n)$ are shown to 
survive quantization, unspoiled by anomalies; 
these theories are therefore integrable at the 
quantum level. Local, higher-spin, conserved charges are also shown 
to survive quantization in the $SO(2n)/SO(n){\times}SO(n)$ models.}} 

\vspace{1cm}
Classical, two-dimensional sigma models on 
compact symmetric spaces $G/H$ are integrable by virtue of 
conserved quantities which can arise as integrals of local 
or non-local functions of the underlying fields  
(the accounts in \cite{EF}-\cite{Evans00} contain 
references to the extensive literature).
Since these models are asymptotically free and strongly coupled in the 
infrared, their quantum properties are not straightforward to determine.
Nevertheless, following L\"uscher \cite{Luscher}, 
Abdalla, Forger and Gomes showed \cite{AFG} that, in a $G/H$ sigma model
with $H$ simple\footnote{Here, and throughout
this paper, we shall use `simple' to
mean that the corresponding Lie algebra has no non-trivial ideals. 
Hence $U(1)$ is simple in our terminology, in addition to the usual
non-abelian simple groups of the Cartan-Killing 
classification \cite{Helgason}.}, the first conserved non-local charge 
survives quantization 
(after an appropriate renormalization \cite{Luscher,AFG,Bernard}), 
which suffices to ensure 
quantum integrability of the theory.
By contrast, calculations using the $1/N$ expansion reveal anomalies
that spoil the conservation of the quantum non-local charges
in the $CP^{N-1}=SU(N)/SU(N{-}1){\times}U(1)$ models for $N > 2$,
and in the wider class of theories based on the complex Grassmannians
$SU(N)/SU(n){\times}SU(N{-}n){\times}U(1)$ for $N>n>1$ \cite{AAG}. 

It was long suspected, therefore, that the $G/H$ sigma models were quantum 
integrable \emph{only} for $H$ simple.
So it was something of a surprise when exact S-Matrices were proposed 
for the family of models based on $SO(2n)/SO(n){\times}SO(n)$, which 
were then shown to pass stringent tests using the Thermodynamic Bethe 
Ansatz (TBA) \cite{Fendley:2000,Fendley:2001}. This was followed by 
the construction of S-matrices for the models with target spaces 
$Sp(2n)/Sp(n){\times}Sp(n)$ \cite{Babichenko}, which were again shown
to be consistent with TBA calculations.

In this letter we reconcile these recent S-matrix results with the
previous, well-known approach of \cite{AFG}, by showing that the latter
techniques can, in fact, be used to show that the first non-local 
charge \emph{does} survive quantization, unspoiled by anomalies,
in the sigma-models with target spaces $SO(2n)/SO(n){\times}SO(n)$
(for $n \geq 3$) 
and $Sp(2n)/Sp(n){\times}Sp(n)$ (for $n \geq 1$). We also argue that these 
techniques cannot be extended to any other new classes of models, at 
least in any obvious way: the non-local charge is protected from anomalies
only if $H$ is simple or if the target space belongs to 
one of these two additional families of Grassmannians. 
As a supplement to our discussion, we
will show at the end of the paper how the quantum integrability of the
$SO(2n)/SO(n){\times}SO(n)$ models can also be established using a
\emph{local} conservation law.

We begin by summarizing the construction of the $G/H$
sigma model \cite{EF,EFlocal}. Let \be \g=\h \oplus \m \ee
be the decomposition of the Lie algebra $\g$ of the compact group $G$ 
into the Lie algebra $\h$ of $H$ and its 
orthogonal complement $\m$; the condition for $G/H$ to be a symmetric
space is 
\be [\h,\h]\subset\h,\quad[\h,\m]\subset\m,\quad[\m,\m]\subset\h.
\label{symm}\ee
The sigma model can be formulated using  
fields $g(x^\mu) \in G$ and $A_\mu(x^\mu)\in\h$ which are subject to
gauge transformations
\be
g(x^\mu)\mapsto g(x^\mu)h(x^\mu)
\, , \qquad
A_\mu \mapsto h^{-1} A_\mu h + h^{-1} \del_\mu h
\label{gauge}\ee
for any $h(x^\mu) \in H$, thus ensuring that the physical degrees of
freedom belong to $G/H$. The fields also transform under a 
global $G$ symmetry 
\be
g(x^\mu)\mapsto U g(x^\mu)
\, , \qquad
A_\mu \mapsto A_\mu 
\label{global}\ee
for any $U \in G$. The lagrangian for the theory, which is 
invariant under each of these symmetries, is
\be {\cal L} =
-{1 \over 2 \lambda} \tr\left( k_\mu k^\mu \right) =
-{1 \over 2 \lambda} \tr\left( j_\mu j^\mu \right)
\label{lagrangian}\ee 
where we use the covariant derivative $D_\mu g \equiv \del_\mu g - g A_\mu$
to define the related, $\g$-valued currents
\bea k_\mu & \equiv & g^{-1} D_\mu g = g^{-1} \del_\mu g - A_\mu \\
j_\mu &\equiv& -(D_\mu g)g^{-1}=-g k_\mu g^{-1} \, . \label{jcurr}
\eea 
Note that $k_\mu$ is gauge-covariant, transforming as
$k_\mu \mapsto h^{-1} k_\mu h$ under (\ref{gauge}),
but it is invariant under (\ref{global}); its covariant derivative is 
$D_\mu k_\nu \equiv \del_\mu k_\nu + [ A_\mu , k_\nu]$.
In contrast, $j_\mu$ is
gauge-invariant, but transforms in the adjoint representation of $G$;
it is the Noether current for the global symmetry (\ref{global}).

The gauge field $A_\mu$ is non-dynamical and the effect of varying it
in the lagrangian is to impose the constraint 
$k_\mu \in \m$. The equation of motion obtained by varying $g$ can
be written in terms of either current:
\be
D_\mu k^\mu = \del_\mu k^\mu + [ A_\mu , k^\mu ]= 0  \qquad \iff \qquad
\del_\mu j^\mu = 0 \, .
\ee
It is now that the symmetric space condition (\ref{symm}) enters
crucially for the first time, because it implies, 
in conjunction with $k_\mu \in \m$, the identities
\bea 0 & = & D_\mu k_\nu - D_\nu k_\mu  \  \in \ \m \\
F_{\mu \nu} & \equiv & \del_\mu A_\nu - \del_\nu A_\mu + [ A_\mu , A_\nu] 
= - [k_\mu , k_\nu] \ \in \ \h
\eea
Equivalently, we have the 
zero-curvature condition\footnote{This terminology is standard but
potentially confusing. The curvature it refers to is that of $j_\mu$ 
regarded as a connection, and not the curvature $F_{\mu \nu}$ of the
non-dynamical gauge field $A_\mu$.} for the gauge-invariant current:
\be \del_\mu j_\nu - \del_\nu j_\mu + 2\left[ j_\mu , j_\nu \right] =
0.
\ee
This, together with the conservation of $j_\mu$, is sufficient to 
show that the $\g$-valued non-local charge
\be Q(t) = \int \! dx \, j_1(t,x) \,
+ \int \! \! \int \! dx dy \, \theta(x{-}y) 
\left[ \, j_0(t,x), \, j_0(t,y) \, \right] \label{Q}\ee
is conserved, which guarantees the integrability of the 
model at the classical level.

The crucial question to be settled in the quantized theory is whether the
definition and conservation of the non-local charge, and hence the
integrability of the theory, can be maintained.
A potential problem arises from the second term in (\ref{Q}): it contains
products of operators at the same spacetime point, and therefore 
entails a careful regularization and renormalization of $Q$.
The approach of \cite{Luscher,AFG,Bernard} is to use 
point-splitting regularization and consider the short-distance
behaviour of the bracket expressed as an operator product expansion (OPE)
\be \left[ \, j_\mu (t,x{+}\eps), \, j_\nu(t,x{-}\eps) \, \right] \,
\sim \,
\sum_k C^{(k)}_{\mu \nu}(\eps) \, Y^{(k)}(t,x)\, .\ee  
Here $\{ Y^{(k)}(t,x)\}$ is a complete set of local operators
of canonical dimension at most two and
$C^{(k)}_{\mu \nu}(\eps)$ are c-number-valued functions which can be 
singular as $\epsilon \rightarrow 0$. We include in the OPE all terms 
which are divergent or non-zero in the limit $\epsilon \rightarrow 0$.

The operator product expansion must, however, transform correctly
under all of the symmetries of the theory. The left-hand side 
transforms under the adjoint
action of the global symmetry $G$ in (\ref{global}),
and is invariant under gauge transformations (\ref{gauge}).
Thus each operator $Y$
in the expansion on the right-hand side must also transform in this way.
But any such operator can be written $Y = g X g^{-1}$, where $X$ is 
invariant under the global $G$ symmetry and instead
transforms covariantly as $X\mapsto h^{-1} X h$ under gauge transformations.
The task is therefore to determine all operators $X$ of this type 
with mass dimension two or less.

There is a unique gauge-covariant operator of dimension one, namely 
the current $k_\mu$, and there are two obvious candidates 
with the correct transformation properties and dimension two,
$D_\mu k_\nu$ and the curvature $F_{\mu \nu}$ of the connection
$A_\mu$. Let us assume for the moment that these are the only
operators that appear. Then, since $F_{\mu \nu}$ is antisymmetric and 
$D_\mu k_\nu$ is symmetric, the OPE takes the form
\bea \left[ \, j_\mu (t,x{+}\eps), \, j_\nu(t,x{-}\eps) \, \right] 
&\sim& C_{\mu \nu}^\rho(\eps) \, g k_\rho g^{-1}
+ C_{\mu \nu}^{\rho \sigma}(\eps) \, 
g\left(D_\rho k_\sigma  +F_{\rho \sigma}\right)g^{-1}\label{ope1}\\
&=& -C_{\mu \nu}^\rho(\eps) \, j_\rho 
- C_{\mu \nu}^{\rho \sigma}(\eps) \, \del_\sigma j_\rho,\label{ope2}\eea
where the coefficient functions $C_{\mu \nu}^\rho (\epsilon)$ 
and $C_{\mu \nu}^{\rho \sigma} (\epsilon)$ are respectively linearly
and logarithmically divergent as $\epsilon \rightarrow 0$ 
(we have suppressed the common spacetime 
argument $(t,x)$ for all the operators on the right-hand side).
The resulting expression (\ref{ope2}) depends 
only on $j_\mu$ and its derivatives, and this is sufficient
\cite{Luscher,AFG,Bernard} to show that the charge $Q$ can be properly defined 
in the quantum theory and that it is conserved. For completeness, 
we give a sketch of the arguments in an appendix.

The key assumption above, that $k_\mu$, $D_\mu k_\nu$ 
and $F_{\mu \nu}$ are the only gauge-covariant terms that can
appear in the OPE, is certainly valid 
if each of these operators transforms in an irreducible
representation of $H$, because there are no other local,
gauge-covariant quantities of the correct dimensions 
that can be constructed from the constituent fields. 
Both $k_\mu$ and $D_\mu k_\nu$ take values in $\m$,
which \emph{always} carries an irreducible representation of $H$ for a
compact symmetric space $G/H$ with $G$ simple \cite{Helgason}. 
But $F_{\mu \nu}$ is valued in $\h$, which
carries the adjoint representation of $H$, and this is irreducible if and 
only if $H$ is simple\footnote{See footnote 1.}.
If $H= H_1 {\times} H_2 {\times} \ldots {\times} H_r$, with a corresponding
decomposition of the Lie algebra 
$\h = \h_1 \oplus \h_2 \oplus \ldots \oplus \h_r $, the total 
curvature can be decomposed into irreducible components
$F_{\mu \nu} = F_{\mu \nu}^{(1)} + F_{\mu \nu}^{(2)} + \ldots + 
F_{\mu \nu}^{(r)} $ where each $F_{\mu \nu}^{(i)} \in \frak{h}_i$
transforms non-trivially only under $H_i$.
In these circumstances our key assumption breaks down because the term 
$ C_{\mu \nu}^{\rho \sigma} g F_{\rho \sigma} g^{-1}$ 
in the OPE (\ref{ope1}) must then be replaced by 
\be
C^{(1) \, \rho \sigma}_{\phantom{x} \mu \nu} g F_{\rho \sigma}^{(1)} g^{-1} +
C^{(2) \, \rho \sigma}_{\phantom{x} \mu \nu} g F_{\rho \sigma}^{(2)} g^{-1} +
\ldots +
C^{(r) \, \rho \sigma}_{\phantom{x} \mu \nu} g F_{\rho \sigma}^{(r)}
g^{-1} \ .
\label{ope3}\ee
The coefficient functions $C^{(i) \, \rho \sigma}_{\phantom{x} \mu \nu}$
are unrelated to one another 
in general, and so it will not generally be possible to re-express
this OPE solely in terms of $j_\mu$ and its derivatives as in (\ref{ope2}).
The conclusion of \cite{AFG} was thus that one should expect anomalies 
to spoil the conservation of $Q$ whenever $H$ is not simple.

But consider now the target spaces $SO(2n)/SO(n){\times}SO(n)$. 
This family is clearly rather special in that, while 
$\h$ is not simple, it is the direct sum
\be \h=\frak{so}(n)_1 \oplus \frak{so}(n)_2 \ee 
of two identical subalgebras, and these subalgebras are simple, 
provided $n \neq 4$. Since neither of the
subalgebras is in any way preferred, 
it is then natural to expect that $g F_{\mu \nu}^{(1)} g^{-1}$ and 
$g F_{\mu \nu}^{(2)} g^{-1}$ should have the \emph{same} 
coefficient in the OPE, in which case 
the usual argument for quantum conservation 
of the non-local charge would still hold. 
The way to formulate this idea precisely is to
show that there is a discrete symmetry $\tau$ of the 
target space $SO(2n)/SO(n){\times}SO(n)$ which exchanges the roles of 
the two factors in the denominator.

The existence of the discrete symmetry $\tau$ is perhaps 
most easily understood by recalling that points on the Grassmannian
$SO(N)/SO(n){\times}SO(N{-}n)$ 
can be identified with $n$-dimensional subspaces in 
$N$-dimensional Euclidean space. 
The factors in the denominator are the linear isometry groups of such an
$n$-dimensional subspace and its orthogonal complement.
The special feature which arises when $N=2n$ is simply that the orthogonal 
complement to an $n$-dimensional subspace is itself $n$-dimensional,
and so $\tau$ can be defined as the map which exchanges these subspaces. 
This is an isometry of the Grassmannian and, therefore, a symmetry of
the sigma-model. 

To express $\tau$ in more concrete terms, consider the following 
block forms for general elements of 
$\frak{so}(2n)$ and its subalgebra $\frak{so}(n)_1 \oplus \frak{so}(n)_2$:
\be
\bmx P & R \\ -{\widetilde R} & Q \emx \in \frak{so} (2n) \, , \qquad
\bmx P & 0 \\  0      & 0 \emx \in \frak{so} (n)_1 \, , \qquad
\bmx 0 & 0 \\  0      & Q \emx \in \frak{so} (n)_2 
\label{block}\ee
($P$, $Q$ and $R$ are $n{\times}n$ real matrices with $P$ and $Q$ both 
antisymmetric and a tilde denotes a transpose). Let 
\be T = \bmx 0 & 1 \\ -1 & 0 \emx \in SO(2n)\label{Tdef}\ee
and consider the inner automorphism of $\frak{so}(2n)$ defined by
\be \tau: \quad \bmx P & R \\ -{\widetilde R} & Q \emx \ \ \mapsto \ \ 
T^{-1} \! \bmx P & R \\ -{\widetilde R} & Q \emx T
       \ = \ \bmx Q & {\widetilde R} \\ -R & P \emx \label{taumat}\ee
which evidently maps $\m \rightarrow \m$ and $\h \rightarrow \h$ 
in such a way that the entries of the $\frak{so}(n)_1$ and
$\frak{so}(n)_2$ subalgebras are interchanged.
From this, we define a transformation on sigma model fields 
\be \tau: \quad 
g\mapsto gT \, , \quad A_\mu \mapsto T^{-1} A_\mu T\label{taufields}\ee
which leaves the Lagrangian (\ref{lagrangian}) invariant;
note also the behaviour of the currents and field strength\footnote{In
the definition (\ref{taufields}) we have chosen to combine the
automorphism acting on $g$ with left multiplication by $T$ 
so as to ensure that the current $j_\mu$ is
invariant, rather than covariant, under the symmetry. 
This is helpful, but not essential, for the arguments that follow.}
\be \tau: \quad 
j_\mu \mapsto j_\mu \, , \quad F_{\mu \nu} \mapsto T^{-1} F_{\mu \nu}T
\, .
\label{taucurv}\ee

Now, terms in the current commutator OPE must be 
invariant under $\tau$, because the currents themselves are. 
The two irreducible components of the curvature (for $n \neq 4$) 
have the block forms 
\be
F_{\mu \nu}^{(1)} =  
\bmx P_{\mu \nu} & 0 \\  0      & 0 \emx \in \frak{so} (n)_1 
\ , \qquad
F_{\mu \nu}^{(2)} = \bmx 0 & 0 \\  0      & Q_{\mu \nu} 
\emx \in \frak{so} (n)_2 \ 
\ee
and it follows from (\ref{taumat}) and
(\ref{taucurv}) that the action of $\tau$ on $F^{(1)}_{\mu \nu}$
and $F^{(2)}_{\mu \nu}$ is to exchange $P_{\mu \nu} \leftrightarrow 
Q_{\mu \nu}$. The combinations 
$g F_{\mu \nu} g^{-1} =  
g (F^{(1)}_{\mu \nu} + F^{(2)}_{\mu \nu} )g^{-1}$
and $g (F^{(1)}_{\mu \nu} - F^{(2)}_{\mu \nu} )g^{-1}$
are clearly even and odd, respectively, under $\tau$
and the OPE must therefore take the form (\ref{ope1}), as claimed.
In essence, the symmetry which constrains the OPE here
is actually the semi-direct product 
\be \mathbb{Z}_2^{(\tau)} \ltimes SO(n)_1\times SO(n)_2 \, .\ee 
Although $F_{\mu \nu}\in\frak{so}(n)_1\oplus\frak{so}(n)_2$ carries a 
reducible representation of $SO(n)_1{\times}SO(n)_2$, 
it carries an \emph{irreducible} representation of this
larger group and so no decomposition of $F_{\mu \nu}$ is
allowed in the OPE.

The existence of the discrete symmetry $\tau$ has thus  
enabled us to extend the approach of \cite{AFG,Luscher} and 
deduce that the quantum $SO(2n)/SO(n){\times}SO(n)$ sigma models possess
conserved non-local charges, ensuring quantum integrability,
for $n \neq 4$. The model with $n = 4$ is also quantum integrable, 
and for exactly similar reasons, but this deserves some additional 
explanation.

The denominator of the symmetric space $SO(8)/SO(4){\times}SO(4)$ 
involves four simple factors rather than two:
\be \h \, = \, \frak{so}(4)_1 \oplus \frak{so}(4)_2 
\, \cong \, 
(\, \frak{su}(2) \oplus \frak{su}(2) \, )  \, \oplus \, 
( \, \frak{su}(2) \oplus \frak{su}(2) \, )  \ .
\label{denom}\ee 
There are then four irreducible curvature components appearing in 
(\ref{ope3}), but discrete symmetries of the sigma model again force
all of the OPE coefficient functions to be equal, as required.
This is actually a consequence of our original symmetry
$\tau$, which exchanges $\frak{so}(4)_1$ and $\frak{so}(4)_2$, 
and just one additional symmetry $\tau'$, constructed 
so as to interchange the $\frak{su}(2)$ subalgebras \emph{within} each copy of
$\frak{so}(4)$. For a single copy of $\frak{so}(4)$,
the $\frak{su}(2)$ subalgebras can be exchanged by conjugating by 
a $4{\times}4$ matrix such as $L = {\rm diag} (1, -1, -1, -1)$ 
(there are many possible choices for $L$; any two differ by an element
of $SO(4)$).
We can therefore define the desired symmetry $\tau'$ by replacing 
$T$ with $T'$ in (\ref{taumat}) and (\ref{taufields}), where
\be T' = \bmx L & 0 \\ 0 & L \emx \in SO(8)\label{Tpdef} \ .\ee
There are other, more exotic discrete symmetries of the
$SO(8)/SO(4){\times}SO(4)$ model
which arise from outer automorphisms of $\frak{so}(8)$ \cite{Helgason} 
and which permute the four $\frak{su}(2)$ subalgebras in (\ref{denom})
in any desired way, but these need not concern us here.

Turning now to the 
$Sp(2n)/Sp(n){\times}Sp(n)$ sigma models with $n \geq 1$,
we can apply almost identical arguments to those used above for the 
real Grassmannians.
Block forms for general elements of $\frak{sp}(2n)$ and its subalgebra 
$\frak{sp}(n)_1 \oplus
\frak{sp}(n)_2$ can be obtained from (\ref{block}) by taking  
$P$, $Q$, $R$ to be $2n{\times}2n$ complex matrices, with 
$P$ and $Q$ antihermitian and the tilde in (\ref{block}) 
denoting hermitian conjugation; these matrices must also satisfy 
\be 
PJ - JP^* = QJ - JQ^* = RJ - JR^* = 0
\ee
where $J$ is a $2n{\times}2n$ symplectic structure (a real,
antisymmetric matrix with $J^2=-1$). The block form for $T$ in (\ref{Tdef})
and the definition of $\tau$ in (\ref{taufields}) are unchanged,
and the reasoning which restricts the form of the OPE and hence
implies the quantum conservation of the non-local charge proceeds just 
as before.

Similar results cannot be expected for 
other compact symmetric spaces $G/H$ with $H$ non-simple, however. 
Our arguments require that $H$ consists of a
product of identical simple subgroups, 
\emph{and} that there is a group of discrete symmetries 
which acts transitively on these factors. 
The first condition holds for the 
families ${SO(2n)}/{SO(n){\times}SO(n)}$ and ${Sp(2n)}/{Sp(n){\times}Sp(n)}$
and for just one other case, namely ${G_2}/{SU(2){\times}SU(2)}$
\cite{Helgason}. 
(Note that the complex Grassmannians 
$SU(2n)/SU(n){\times}SU(n){\times}U(1)$ are ruled out because of the 
extra $U(1)$ factor in the denominator.) 
For the second condition to hold, it
must be possible to introduce $\tau$ as an automorphism of $\frak{g}$
which commutes with the involutive automorphism defining $G/H$,
and which permutes the simple
factors in $\frak{h}$.\footnote{It is 
interesting to note that for $\frak{g} = \frak{so}(2n)$
and $\frak{g} = \frak{sp}(2n)$ the involutive automorphism 
$\tau$ defines the symmetric spaces
$SO(2n)/U(n)$ and $Sp(2n)/U(2n)$ respectively. Commuting, involutive
automorphisms have recently proved useful in classifying integrable
boundary conditions for symmetric space sigma models on the half line
\cite{MY}.
}. 
This is not possible 
for $G_2/SU(2){\times}SU(2)$ because the two $SU(2)$ factors can 
be distinguished: they are embedded inequivalently in $G_2$ 
and they act in different representations on $\m$ \cite{Helgason}.

To conclude our discussion, we will give an alternative demonstration 
of the quantum integrability of the sigma models on 
$SO(2n)/SO(n){\times}SO(n)$, quite independent of the non-local charges that 
have been the subject of the paper so far.
 
The classical integrability of the $G/H$ symmetric space sigma
models can also be understood in terms of higher-spin 
conserved currents that are 
\emph{local} in the fields and are related to $H$-invariant symmetric 
tensors on $\frak{m}$ \cite{Evans00}.
It is usually difficult to draw conclusions about the
survival of such conservation laws at the quantum level, but there 
are some notable exceptions which can be analysed very
simply using an approach due to Goldschmidt and Witten \cite{GW}. 
Their method entails enumerating all possible terms which could
violate a given classical conservation equation, and comparing 
with the number of such terms which can be written as total derivatives,
and whose appearance would therefore constitute a modification of the 
conservation equation, rather than a violation of it. 
Global symmetries in general, and discrete symmetries in particular, 
again play a crucial role.

To carry out such an analysis for the $SO(2n)/SO(n){\times}SO(n)$ 
sigma model it is convenient to reformulate it using a field
$\Phi^{ab} (x^\mu)$
which is a real, symmetric, traceless $2n{\times}2n$ matrix
constrained to satisfy $\Phi^2 = 1$ (this is used in \cite{Fendley:2000}).
The lagrangian for $\Phi$ is free except for the constraint, and 
the equations of motion are easily found (using a Lagrange 
multiplier) to be
\be
\del^\mu \del_\mu \Phi \, + \, \Phi (\del^\mu \Phi)( \del_\mu \Phi) \,
= \, 0 \ .
\ee
There are no gauge fields in this formulation 
and the $SO(2n)$ global symmetry (\ref{global}) 
acts by $ \Phi \mapsto U \Phi U^T$. Actually, the symmetry extends to $O(2n)$
by including a transformation
\be \mu : \ \ \Phi \mapsto M \Phi M^T, \qquad  M M^T = 1 \ , \qquad
\det M = -1 \ .
\ee
The discrete symmetry (\ref{taufields}) is now simply 
\be 
\tau: \ \ \Phi \mapsto -\Phi \ .
\ee
(The relation of this new formulation to our previous
description of the model is revealed by writing $\Phi = g N g^T$, 
where $N= {\rm diag} (1,-1)$ in the basis (\ref{block}) and 
$g \in SO(2n)$ with a redundancy
$g \mapsto gh$, $h \in SO(n){\times}SO(n)$.)

In this new notation, the Noether currents (\ref{jcurr}) are antisymmetric
matrices $j_\mu^{ab}$, whose definition and conservation may be
written
\be
j_\mu \, = \, {\textstyle {1 \over 2}} \, \Phi \, \del_\mu \Phi \, , 
\qquad
\del_\mp j_\pm \, = \pm [ j_+  , j_- ]
\ee
(light-cone components for vectors in Minkowski space are defined 
by $u_\pm = u_0 \pm u_1$).
A local, classically-conserved quantity can be constructed from 
$j_\mu$ using any symmetric invariant tensor, but we shall concentrate here on
the Pfaffian for $SO(2n)$ which yields the conservation
law\footnote{It is important to check that this higher-spin current 
does not vanish \emph{identically}. The analysis of \cite{Evans00}
ensures this: there is a non-vanishing invariant on 
$SO(2n)/SO(n){\times}SO(n)$ inherited from the Pfaffian 
for $SO(2n)$, but this is not true for other real Grassmannians.
Note also that conserved quantities 
in \cite{Evans00} are written in terms of the gauge-covariant currents,
$k_\mu$ rather than $j_\mu$ in our present notation.}
\be \del_- ( \, \varepsilon_{a_1 b_1 a_2 b_2 \ldots a_n b_n} \, 
j_+^{a_1 b_1} j_+^{a_2 b_2} 
\dots j_+^{a_n b_n} \, ) = 0\label{pfaff}\ee
This higher-spin current is clearly even under $\tau$, 
but it is odd under $\mu$, since the Pfaffian transforms with a factor 
$\det M = -1$, and it is this which proves particularly useful in restricting 
the possible quantum corrections.

We now consider all 
local operators constructed from $\Phi$ and its derivatives 
whose symmetry properties allow them to 
appear as quantum corrections on the right-hand side of (\ref{pfaff}). 
To form an $SO(2n)$ invariant, the indices on all fields $\Phi^{ab}$,
$\del_\pm \Phi^{ab}$, and higher derivatives,
must be contracted with each other (using $\delta_{ab}$) or with 
$\varepsilon_{a_1 a_2 \ldots a_{2n}}$.
An $\varepsilon$-tensor is essential here, however, because without one we 
can construct only traces of products of matrices, which will all be even 
under $\mu$. The antisymmetry of the $\varepsilon$-tensor then 
severely limits which products of matrices can be contracted with it, and 
we have the freedom to move matrices around within a 
product by using identities such as $\Phi (\del_\mu \Phi) = -
(\del_\mu \Phi) \Phi$ which are consequences of the constraint $\Phi^2=1$. 
Finally, the symmetry $\tau$ is important in
restricting the total number of fields $\Phi$ 
(including derivatives) to be even. 
Taking all these facts into account, we find that, 
up to terms which vanish on using the equations of motion, any 
quantum modification of (\ref{pfaff}) is proportional to
\be \del_+ ( \, \varepsilon_{a_1 b_1 a_2 b_2 \dots a_n b_n} \, 
j_-^{a_1 b_1} j_+^{a_2 b_2} \dots j_+^{a_n b_n} \, )  \ . \ee
Because this is a derivative, the conservation law is guaranteed to
survive at the quantum level, albeit in a modified form.
\\

{\bf Appendix: the quantum non-local charge and its conservation}

The definition of the quantum non-local charge given in 
\cite{Luscher,AFG,Bernard} is
\bea Q_\delta(t) &=& Z(\delta) \! \int \! dx \, j_1(t,x) \ + \
\int \! \int  \! dx dy \, \theta(x{-}y{-}\delta) \, \left[ \, j_0(t,x), \,
  j_0(t,y) \,  \right]\nn\\
&=& \int_{-\infty}^{\infty} dx \left( \, Z(\delta) j_1(t,x) \,  
+ \, \int_{\delta}^{\infty} \! d\eps \, \left[ \, j_0(t,x), \,
  j_0(t,x{-}\eps) \, \right]\right)
\label{Qd},\eea
where we must show that the cut-off can be removed, $\delta
\rightarrow 0$, after choosing the renormalization factor  
$Z(\delta)$ so as to cancel the divergence arising from the 
commutator term. Notice that terms in the integrand
which are logarithmically-divergent or finite as $\eps \rightarrow 0$ will
\emph{not} give rise to divergences in the integral as $\delta \rightarrow
0$. But the form of the remaining, linearly-divergent term in the OPE  
is fixed by two-dimensional spacetime symmetries (Lorentz, parity
and time-reversal invariance) to be of the form
\be \left[ \, j_0(t,x), \, j_0(t,x{-}\eps) \, \right] \, \sim \,
C(\eps) \, j_1(t,x)
\ .\label{opea}\ee
It follows that the charge is well-defined as $\delta\rightarrow 0$ 
provided $Z'(\delta)= C(\delta)$. 

Consider now an expression for the time derivative of the charge:
\be \del_0 Q_\delta(t) 
       = \int_{-\infty}^{\infty} dx \left\{  \, \,  Z(\delta) \del_0 j_1(t,x) 
\, + \, \left[ \, j_0(t,x), \, j_1(t,x{-}\delta)+j_1(t,x{+}\delta) \, 
\right] \, \, \right\} \, ,
\label{Qdot}\ee
which is obtained on using conservation of $j_\mu$, 
a shift in an integration variable, and the relation
\be \int_{-\infty}^{\infty} 
dx \, \del_0 \left[ \, j_0(t,x), \, j_0(t,x{-}\delta) \, \right] = 
-\int_{-\infty}^{\infty} dx \frac{\del}{\del{\delta}}\left[ \,
  j_0(t,x), \, j_1(t,x{-}\delta) + j_1(t,x{+}\delta) \,  \right] \, \,
. \label{trick}\ee
The right-hand side of (\ref{Qdot}) must be finite 
as $\delta\rightarrow 0$ if $Z'(\delta) = C(\delta)$ (since 
we know $Q$ itself is well-defined in this limit) and, indeed,
the identity (\ref{trick}) can be used once more to relate the 
singular part of the OPE for the current commutator 
\be\left[ \, j_0(t,x), \, j_1(t,x{-}\delta) \, + \, j_1(t,x{+}\delta) \,
\right]
\label{currcom}\ee
to the OPE in (\ref{opea}). The conclusion is that, up to space
derivatives, the singular part of (\ref{currcom}) must be of the form 
$W(\delta) \, \del_0 j_1(t,x)$, where $W'(\delta) = C(\delta)$. 
Thus (\ref{Qdot}) is well-defined 
for $Z(\delta) = W(\delta) + a$, with $a$ any constant.

Finally, to determine if the charge is conserved we must examine the
finite, $\delta$-independent terms in the OPE
for (\ref{currcom}). If (\ref{ope2}) holds, the contribution 
to (\ref{Qdot}) is proportional to the integral 
over space of $\del_0 j_1$, and this can be cancelled by 
choosing the constant $a$ appropriately.
If (\ref{ope2}) does not hold, however, then we will in general 
have contributions from the curvature components 
$F^{(i)}_{\mu \nu}$ which cannot be cancelled by any choice of
$Z(\delta)$, and the non-local charge will not be conserved. The
anomalous contributions found for the complex Grassmannians 
in \cite{AAG} are exactly of this type. 
\\

{\bf Acknowledgments}:
The research of JME is supported in part by Gonville and Caius
College, Cambridge. CASY is supported by a PPARC studentship. DK is
grateful for a NSF graduate research fellowship and a Columbia
University Euretta J.~Kellett fellowship.
\pagebreak

{\small
\begin{thebibliography}{99}
\raggedright
%
\bibitem{EF}
H. Eichenherr and M. Forger, {\em On the dual symmetry of the
non-linear sigma models}, Nucl.~Phys.~{\bf B155} (1978) 381; {\em
More about non-linear sigma models on symmetric spaces},
Nucl.~Phys.~{\bf B164} (1980) 528, Erratum-ibid.~{\bf B282}
(1987) 745.

\bibitem{EFlocal} H. Eichenherr and M. Forger, 
{\em Higher local conservation laws for nonlinear sigma models on symmetric
spaces}, Commun.\ Math.\ Phys.\  {\bf 82} (1981) 227.

\bibitem{MacKay92}
N.J.~MacKay,
{\em On the classical origins of Yangian symmetry in integrable field theory,}
Phys.\ Lett.\ {\bf B281} (1992) 90,
Erratum-ibid.\ {\bf B308} (1993) 444.
%%CITATION = PHLTA,B281,90;%%

\bibitem{PCM}
J.M.~Evans, M.~Hassan, N.J.~MacKay and A.J.~Mountain, 
{\em Local conserved charges in principal chiral models},
Nucl.~Phys.~{\bf B561} (1999) 385 [arXiv:hep-th/9902008];
\hfill \break
J.M.~Evans, {\em Integrable sigma-models and Drinfeld-Sokolov
hierarchies}, Nucl.~Phys.~{\bf B608} (2001) 591 [arXiv:hep-th/0101231].

\bibitem{Evans00}
J.M.~Evans and A.J.~Mountain, {\em Commuting charges and symmetric
spaces}, Phys.~Lett.~{\bf B483} (2000) 290 [arXiv:hep-th/0003264].

\bibitem{Luscher}
M.~L\"uscher,
{\em Quantum nonlocal charges and absence of particle production in the
two-dimensional nonlinear sigma model},
Nucl.\ Phys.\ {\bf B135} (1978) 1.
%%CITATION = NUPHA,B135,1;%%

\bibitem{AFG}
E.~Abdalla, M.~Forger and M.~Gomes,
{\em On the origin of anomalies in the quantum nonlocal charge for the
generalized nonlinear sigma models,}
Nucl.\ Phys.\ {\bf B210} (1982) 181.
%%CITATION = NUPHA,B210,181;%%

\bibitem{Bernard}
D.~Bernard, {\em Hidden Yangians in 2D massive current algebras},
Commun.~Math.~Phys.~{\bf 137} (1991) 191.

%\bibitem{Alvarez}
%O.~Alvarez, 
%{\em 1/N paper}, Phys.~Rev.~{\bf D17} (1978) 1123.

\bibitem{AAG}
E.~Abdalla, M.C.B.~Abdalla and M.~Gomes,
{\em Anomaly in the nonlocal quantum charge of the $CP^{N-1}$ model}, 
Phys.\ Rev.\ D {\bf 23} (1981) 1800;
\hfill \break
E.~Abdalla, M.~Forger and A.~Lima Santos,
{\em Non-local charges for non-linear sigma models on Grassmann
manifolds}, Nucl.~Phys.~{\bf B256} (1985) 145.
%%CITATION = PHRVA,D23,1800;%%

\bibitem{Fendley:2000}
P.~Fendley,
{\em Integrable sigma models with $\theta = \pi$},
Phys.\ Rev.\ B {\bf 63} (2001) 104429
[arXiv:cond-mat/0008372].
%%CITATION = COND-MAT 0008372;%%

\bibitem{Fendley:2001}
P.~Fendley,
{\em Integrable sigma models and perturbed coset models},
JHEP {\bf 0105} (2001) 050
[arXiv:hep-th/0101034].
%%CITATION = HEP-TH 0101034;%%

\bibitem{Babichenko}
A.~Babichenko,
{\em Quantum integrability of sigma models on AII and CII symmetric spaces},
Phys.\ Lett.\ {\bf B554} (2003) 96
[arXiv:hep-th/0211114].
%%CITATION = HEP-TH 0211114;%%

\bibitem{Helgason}
S.~Helgason, 
{\em Differential geometry, Lie groups, and symmetric spaces}, 
Academic Press, New York, 1978.

\bibitem{GW}
Y.Y.~Goldschmidt and E.~Witten,
{\em Conservation laws in some two-dimensional models},
Phys.\ Lett.\ {\bf B91} (1980) 392.
%%CITATION = PHLTA,B91,392;%%

\bibitem{MY}
N.J.~MacKay and C.A.S.~Young,
{\em Classically integrable boundary conditions for symmetric-space 
sigma models}, Phys.~Lett.~B, to appear [arXiv:hep-th/0402182].

\end{thebibliography}
}
 \end{document}